\documentstyle[11pt]{article}
\newlength{\bredde}
\def\slash#1{\settowidth{\bredde}{$#1$}\ifmmode\,\raisebox{.15ex}{/}
\hspace*{-\bredde} #1\else$\,\raisebox{.15ex}{/}\hspace*{-\bredde} #1$\fi}
\newcommand{\beq}{\begin{equation}}
\newcommand{\eeq}{\end{equation}}
\newcommand{\noi}{\vspace{12pt}\noindent}
\newcommand{\lG}{\raise.3ex\hbox{$\stackrel{\leftarrow}{G}$}}
\newcommand{\lU}{\raise.3ex\hbox{$\stackrel{\leftarrow}{U}$}}
\newcommand{\lP}{\raise.3ex\hbox{$\stackrel{\leftarrow}{{\cal P}}$}}
\newcommand{\leta}{\raise.3ex\hbox{$\stackrel{\leftarrow}{\eta}$}}
\newcommand{\lOmega}{\raise.3ex\hbox{$\stackrel{\leftarrow}{\Omega}$}}
\newcommand{\ldr}{\raise.3ex\hbox{$\stackrel{\leftarrow}{\delta^r}$}}

\def\beqn{\begin{eqnarray}}
\def\eeqn{\end{eqnarray}}

\def\gtwid{\raise.3ex\hbox{$>$\kern-.75em\lower1ex\hbox{$\sim$}}}
\def\ltwid{\raise.3ex\hbox{$<$\kern-.75em\lower1ex\hbox{$\sim$}}}

\def\la{\lambda}

\begin{document}
\topmargin -1.4cm
\oddsidemargin -0.8cm
\evensidemargin -0.8cm
\title{\Large{{\bf Dirac Operator Spectra from Finite-Volume
Partition Functions}}}

\vspace{1.5cm}

\author{~\\~\\
{\sc Poul H. Damgaard}\\
The Niels Bohr Institute\\ Blegdamsvej 17\\ DK-2100 Copenhagen\\
Denmark}
 
\maketitle
\vfill
\begin{abstract}Based on the relation to random matrix theory, exact 
expressions for all microscopic spectral correlators of the Dirac operator 
can be computed from finite-volume partition functions. This is illustrated
for the case of $SU(N_c)$ gauge theories with $N_c\geq 3$ and $N_f$
fermions in the fundamental representation. 
\end{abstract}
\vfill
\begin{flushleft}
NBI-HE-97-60 \\
hep-th/9711110
\end{flushleft}
\thispagestyle{empty}
\newpage

\setcounter{page}{1}

\noi
Central to the understanding of chiral symmetry breaking in gauge theories
is the spectral density $\rho(\la)$ of the Dirac operator. By the Banks-Casher
relation $\rho(0) \!=\! N\Sigma/\pi~,~\Sigma \!\equiv\! 
\langle\bar{\psi}\psi\rangle$, this spectral density, when evaluated at
the origin, is an order parameter for chiral symmetry breaking. As is
well-known, this spontaneous symmetry breaking can, 
for the massless theory, only occur if 
one first considers the theory in a finite volume $V$, introduces a global
source (mass) $m$ for $\bar{\psi}\psi$, takes the limit of $V\!\to\!\infty$,
and subsequently the limit $m\!\to\!0$. One therefore needs to know
the spectral density for small masses and large volumes, and then trace how
the eigenvalues of the Dirac operator accumulate towards $\la\!=\! 0$ as
the above limit is taken.

\noi
A few years ago Leutwyler and Smilga \cite{LS}
greatly enhanced our understanding
of this issue by deriving a series of exact spectral sum rules for the
Dirac operator of gauge theories in a range $1/\Lambda_{QCD}
\ll V^{1/4} \ll 1/m_{\pi}$. Here $\Lambda_{QCD}$ is a hadronic scale 
in QCD, $V$ is the space-time volume, and $m_{\pi}$ is the pion 
mass. \footnote{Here ``QCD'' is considered in
the general sense of a $(3\!+\!1)$-dimensional gauge theory with gauge
group $SU(N_c\!\geq\!3)$ and $N_f$ fermion species in the fundamental
representation of this gauge group. (Generalizations to other theories
are considered in \cite{LS} and \cite{SmV}). Whenever we refer to the
finite-volume partition function of QCD in what follows, it is
considered in the above ``mesoscopic'' range of volumes.} In doing this,
Leutwyler and Smilga simultaneously derived exact analytical expressions
for the finite-volume partition functions $Z_{\nu}$ of a variety of 
different theories in sectors of fixed topological charge $\nu$.
Based on a quite remarkable connection to random matrix theory 
\cite{SV,V,VZ}, these exact statements about the Dirac operator spectrum
are now understood in a completely new light. The first step in this
series of events was the observation that the so-called
microscopic spectral density of the theory with massless fermions,
\beq
\rho_S(\zeta) 
~\equiv~ \lim_{N\to\infty} \frac{1}{\Sigma N}\rho\left(
\frac{\zeta}{\Sigma N}\right) ~,\label{rhomassless}
\eeq
can reproduce all QCD spectral sum rules if
computed in a large-$N$ random matrix 
model ensemble (the precise ensemble being uniquely dictated by the
symmetries of the Dirac operator). One key to the understanding of this
surprising fact is the by now proven universality, within random matrix
theory, of all massless microscopic spectral correlators, and in 
particular the
microscopic spectral density itself \cite{ADMN}. An
infinite sequence of spectral sum rules for the massless QCD Dirac operator 
can thus be reproduced by one single function, the microscopic spectral 
density defined above. While this is no proof that the microscopic 
spectral correlators are exact statements about QCD in the above limit,
it is highly suggestive. It has also been put to test in 
a direct Monte Carlo evaluation of the microscopic spectral density
(for $SU(2)$, in the quenched approximation) \cite{BBMSVW}.

\noi
Very recently, the above considerations have been extended to the
case of massive fermions in QCD \cite{DN} (see also \cite{WGW}). 
Here the appropriate limit is double-microscopic:
both eigenvalues $\la$ and masses $m$ must be considered on a scale
of comparable magnification. Both for this case, and for 
QCD with massive fermions in $(2\!+\!1)$ dimensions, the corresponding
double-microscopic spectral correlators have been shown to be universal
in the context of random matrix theory \cite{DN,DN1}. Moreover, the
double-microscopic spectral densities satisfy non-trivial massive 
generalizations of the spectral sum rules \cite{DN,DN1,D}.

\noi
One obvious question is the following.
Given the fact that (double-) microscopic spectral correlators computed
in a certain large-$N$ random matrix theory can be used to ``saturate''
the spectral sum rules derived from finite-volume partition
functions, are these spectral correlators themselves derivable
from the finite-volume partition functions? 
There are already some results which would seem to indicate
that this is not the case. For example, the finite-volume partition 
functions, and hence all spectral sum rules, for $SU(2)$ gauge theory 
with 1 fermion and $SU(N_c\!\geq\!3)$ gauge theory with 1 fermion
are identical. However, their microscopic spectral correlators are
{\em different} \cite{V}.
We shall nevertheless show in this paper that all information about the 
microscopic spectral correlators in a very precise sense {\em is} contained 
in the finite-volume partition functions. How the apparent counterexample
above can be understood will become clear as we proceed.

\noi
To be specific, let us consider the most interesting case, that of
gauge group $SU(N_c), N_c \geq 3$, with $N_f$ fermions in the fundamental 
representation. The coset of spontaneous symmetry breaking is here 
$SU(N_f)_L\times SU(N_f)_R/SU(N_f)$, and the corresponding matrix models 
are those of the chiral unitary ensemble \cite{V}. We begin our 
considerations in the large-$N$ random matrix model language. Here
partition function reads, for the sector of topological charge 
$\nu$, \cite{V}
\beq
\tilde{\cal Z}_{\nu}^{(N_{f})}(m_1,\ldots,m_{N_{f}}) 
~=~ \int\! dW \prod_{f=1}^{N_{f}}{\det}\left(M + im_f\right)~
\exp\left[-\frac{N}{2} {\rm tr}\, V(M^2)\right] ~,
\eeq
where
\beq
M ~=~ \left( \begin{array}{cc}
              0 & W^{\dagger} \\
              W & 0
              \end{array}
      \right) ~.
\eeq
Here $W$ is a rectangular complex matrix of size
$N\times(N\! +\! |\nu|)$. In the large-$N$ limit the space-time volume $V$ 
of QCD is identified with $2N$. We integrate over the Haar measure of $W$.

\noi
In terms of the eigenvalues $\la_i$ of the hermitian matrix $W^{\dagger}W$
the partition function  can be written (ignoring unimportant overall factors)
\beq
\tilde{\cal Z}_{\nu}^{(N_{f})}(m_1,\ldots,m_{N_{f}})  ~=~ 
\prod_f(m_f^{\nu})\int_0^{\infty}\! \prod_{i=1}^N \left(d\lambda_i 
~\la_i^{\nu}~\prod_{f=1}^{N_{f}}(\lambda_i + m_f^2)~
{\rm e}^{-NV(\lambda_i)}\right)\left|{\det}_{ij}
\lambda_j^{i-1}\right|^2 ~.\label{zmatrixeigen}
\eeq
It is of course a symmetric function in $m_f$.
Moreover, one reads off that apart for the overall factor
of $\prod m_f^{\nu}$, the partition function for $N_{f}$ fermions
in the sector of topological charge $\nu$ is equal to the partition 
function of the same $N_{f}$ fermions plus $\nu$ additional massless fermions
of zero mass, in the sector of zero topological charge. For simplicity,
we set $\nu\!=\! 0$ in what follows, so that there are no zero modes.

\noi
A convenient expression for the spectral correlators
of random hermitian matrices (the unitary ensemble) has recently been given 
in \cite{ZJ}. The appropriate generalization to the case of our chiral
unitary ensemble with measure (\ref{zmatrixeigen}), 
where we are interested in the correlators of 
eigenvalues $z_i$ of $M$ rather than those, $\la_i\! =\! z_i^2$, 
of $W^{\dagger}W$, is straightforward. 
The two-point correlator, the kernel, is 
\beq
K^{(N_{f})}_N(z,z';m_1,\ldots,m_{N_{f}}) ~=~ 
e^{-\frac{N}{2}(V(z^2)+V(z'^2))}\sqrt{zz'}
\prod_{f}\sqrt{(z^2+m_f^2)(z'^2+m_f^2)}
\sum_{i=0}^{N-1} P_i(z^2)P_i(z'^2) ~,
\eeq 
where $P_i(z^2)$ are the usual (orthonormal) polynomials associated with 
the above matrix model (see, $e.g.$, ref. \cite{ADMN}). 
They of course depend on all masses $m_f$. The kernel can now be
expressed as a normalized random matrix integral itself:
\begin{eqnarray}
K^{(N_{f})}_N(z,z';m_1,\ldots,m_{N_{f}})\! &=& \!
\frac{e^{-\frac{N}{2}(V(z^2)+V(z'^2))}\sqrt{zz'}
\prod_{f}\sqrt{(z^2+m_f^2)(z'^2+m_f^2)}}{
\tilde{\cal Z}_{0}^{(N_{f})}(m_1,\ldots,m_{N_{f}})} ~\times \cr
&& \!
\int_0^{\infty}\! \prod_{i=1}^{N-1}\!\!\left(\!d\lambda_i 
(\la_i-z^2)(\la_i-z'^2)\!\!\prod_{f=1}^{N_{f}}(\lambda_i + m_f^2)
{\rm e}^{-NV(\lambda_i)}\!\right)\!\left|{\det}_{ij}
\lambda_j^{i-1}\right|^2 ~.
\end{eqnarray}
The last integral is over $(N\!-\!1)$ eigenvalues only. However, in the
large-$N$ limit we shall consider below, this distinction can be ignored.
Thus, in the large-$N$ limit we have
\begin{eqnarray}
K^{(N_{f})}_N(z,z';m_1,\ldots,m_{N_{f}}) &=&
e^{-\frac{N}{2}(V(z^2)+V(z'^2))}\sqrt{zz'}
\prod_{f}\sqrt{(z^2+m_f^2)(z'^2+m_f^2)}\cr
&&\times ~\frac{
\tilde{\cal Z}_{0}^{(N_{f}+2)}(m_1,\ldots,m_{N_{f}},iz,iz')}{
\tilde{\cal Z}_{0}^{(N_{f})}(m_1,\ldots,m_{N_{f}})} ~,
\end{eqnarray}
where the matrix model partition function in the numerator is evaluated
for a theory corresponding to $(N_f\!+\!2)$ fermions, of which two 
have imaginary mass. By means of the usual factorization property, all
higher $n$-point spectral correlation functions are then also explicitly
expressed in terms of the two matrix model partition functions
$\tilde{\cal Z}_{0}^{(N_{f})}$ and $\tilde{\cal Z}_{0}^{(N_{f}+2)}$.
The spectral density corresponds to the two additional (imaginary)
masses being equal:
\beq
\rho^{(N_{f})}(z;m_1,\ldots,m_{N_{f}}) ~=~
\lim_{N\to\infty} K_N^{(N_{f})}(z,z;m_1,\ldots,m_{N_{f}}) ~.
\eeq

\noi
We now consider the double-microscopic limit in which $\zeta \equiv
z N2\pi\rho(0)$ and $\mu_i \equiv m_i N2\pi\rho(0)$ are kept fixed as
$N\!\to\!\infty$. In this limit the pre-factor 
$\exp[-(N/2)(V(z^2)+V(z'^2))]$ becomes replaced by unity. Identifying
$\Sigma = 2\pi\rho(0)$, this is also the limit in which we can compare
with the finite-volume partition function of QCD.

\noi
What is the relation between the finite-volume QCD partition function
${\cal Z}_{0}^{(N_{f})}(\mu_1,\ldots,\mu_{N_{f}})$ and the matrix model 
partition $\tilde{\cal Z}_{0}^{(N_{f})}(\mu_1,\ldots,\mu_{N_{f}})$?
In the mesoscopic scaling region, which corresponds to the double-microscopic
scaling regime of the matrix models, they should just be proportional,
with a proportionality constant that is independent of the masses.
This will then provide us with the sought-for
relation between (double-) microscopic spectral correlators of the
Dirac eigenvalues, and finite-volume partition functions. For the
kernel, the master formula is
\beq
K_S^{(N_{f})}(\zeta,\zeta';\mu_1,\ldots,\mu_{N_{f}}) ~=~
C\sqrt{\zeta\zeta'}\prod_{f}\sqrt{(\zeta^2+\mu_f^2)(\zeta'^2+\mu_f^2)}~\frac{
{\cal Z}_{0}^{(N_{f}+2)}(\mu_1,\ldots,\mu_{N_{f}},i\zeta,i\zeta')}{
{\cal Z}_{0}^{(N_{f})}(\mu_1,\ldots,\mu_{N_{f}})} ~.\label{mf}
\eeq

\noi
The double-microscopic spectral density is thus
\beq
\rho_S^{(N_{f})}(\zeta;\mu_1,\ldots,\mu_{N_{f}}) ~=~
C |\zeta| \prod_{f}(\zeta^2+\mu_f^2)~\frac{
{\cal Z}_{0}^{(N_{f}+2)}(\mu_1,\ldots,\mu_{N_{f}},i\zeta,i\zeta)}{
{\cal Z}_{0}^{(N_{f})}(\mu_1,\ldots,\mu_{N_{f}})} ~,
\eeq
and the double-microscopic $n$-point correlation functions are given by
\beq
\rho_S^{(N_{f})}(\zeta_1,\ldots,\zeta_n;\mu_1,\ldots,\mu_{N_{f}}) ~=~
\det_{a,b} K_S^{(N_{f})}(\zeta_a,\zeta_b;\mu_1,\ldots,\mu_{N_{f}}) ~.
\label{correl}
\eeq
The proportionality constant $C$ is still undetermined, but there are
several ways to fix it. One simple procedure is to use the matching
between the microscopic spectral density 
$\rho_S^{(N_{f})}(\zeta;\mu_1,\ldots,\mu_{N_{f}})$ as $\zeta\!\to\!
\infty$ with the macroscopic spectral density at the origin, or, 
in the conventional normalization \cite{V}, $1/\pi$.

\noi
As a first example illustrating this, consider the microscopic
spectral density $\rho_S^{(0)}(\zeta)$ of quenched QCD, which 
formally corresponds to $N_f\!=\!0$. To find it, we need the
finite-volume QCD partition function for two massive fermions of
degenerate (rescaled) masses $i\mu$. This was evaluated analytically
already in ref. \cite{LS} and found to be, in their normalization,
\beq
{\cal Z}_{0}^{(2)}(i\mu,i\mu) ~=~ I_{0}(i\mu)^2 - 
I_{1}(i\mu)^2 ~,
\eeq
where $I_n(x)$ is the $n$th modified Bessel function. The corresponding
denominator in eq. (\ref{mf}), ${\cal Z}_{0}^{(0)}$, is in this case
just an irrelevant constant which can be set to unity. By means of the
relation $I_n(ix) \!=\! i^nJ_n(x)$ between modified and ordinary Bessel
functions of integer order, this gives
\beq
\rho_S^{(0)}(\zeta) = C~ |\zeta|\left[J_{0}(\zeta)^2
+ J_{1}(\zeta)^2\right] ~.
\eeq
Requiring $\rho_S^{(0)}(\zeta\!\to\!\infty) \!=\! 1/\pi$ yields
$C\!=\! 1/2$, and hence
\beq
\rho_S^{(0)}(\zeta) = \frac{1}{2}|\zeta|\left[J_{0}(\zeta)^2
+ J_{1}(\zeta)^2\right] ~,
\eeq
the known result \cite{V}. No {\em explicit} large-$N$ random matrix
model computation is needed. 
But we can do more than this. The finite-volume QCD partition
function for $N_f$ fermions of arbitrary masses has recently been computed
analytically by Jackson, \c{S}ener and Verbaarschot \cite{JSV}. 
The result is (dropping an irrelevant overall factor):
\beq
Z_{0}^{(N_{f})}(\mu_1,\ldots,\mu_{N_{f}}) ~=~ 
\frac{\det A}{\Delta(\mu^2)} ~,\label{znonequalm}
\eeq
where the $N_f\!\times\! N_f$ matrix $A$ and $\Delta(\mu^2)$ are given by
\beq
A_{ij} ~=~ \mu_i^{j-1}I_{0}^{(j-1)}(\mu_i) ~~,~~~~~~
\Delta(\mu^2) ~=~ \prod_{i<j} (\mu_i^2 - \mu_j^2)~. \label{Aderiv}
\eeq
By repeated differentiation of the Bessel function relation
\beq
xI_n'(x) ~=~ nI_n(x) + xI_{n+1}(x) ~,
\eeq
and by making use of the invariance properties of the determinant, we can
replace the matrix $A$ of (\ref{Aderiv}) by a more convenient expression:
\beq
A_{ij} ~=~ \mu_i^{j-1}I_{j-1}(\mu_i) ~.\label{Aconv}
\eeq

\noi
For the numerator of eq. ({\ref{mf}) we need the $(N_f\!+\!2)\!\times\!
(N_f\!+\!2)$ matrix $A$ with two of the entries being imaginary. 
This means that
\beq
A_{ij} = (-\zeta_i)^{j-1}J_{j-1}(\zeta_i) ~~~~~~~~~~ 
{\mbox{\rm for}}~~i = 1,2 ~,
\eeq
and otherwise (for $i\!\geq\!3$) as in (\ref{Aconv}). Also in the Vandermonde
determinant of (\ref{Aderiv}) it is convenient to separate out explicitly
the terms arising from the imaginary entries. These terms are
$$
-1(\zeta_1^2 - \zeta_2^2)\left(\prod_f(\zeta_1^2+\mu_f^2)(\zeta_2^2+\mu_f^2)
\right) ~,
$$
while the remaining terms are identical to those coming from the
denominator (and they hence cancel). For convenience we now pull out a
factor of (-1) from every second column of the matrix $A$. This yields an
overall factor of $(-1)^{[N_f/2]}$
where $[x]$ denotes the integer part of $x$, and the matrix $A$
is redefined accordingly (we call it $B$ below). 
Putting these pieces together gives
\beq
K_S^{(N_{f})}(\zeta_1,\zeta_2;\mu_1,\ldots,\mu_{N_{f}}) =
C~\frac{(-1)^{[N_f/2]+1}\sqrt{\zeta_1\zeta_2}}{(\zeta_1^2 - \zeta_2^2)
\prod_f\sqrt{(\zeta_1^2+\mu_f^2)(\zeta_2^2+\mu_f^2)}}~\frac{\det B}{\det A} ~,
\label{KS}
\eeq
where the $(N_f\!+\!2)\!\times\!(N_f\!+\!2)$ matrix $B$ is defined by
\begin{eqnarray}
B_{ij} &=& (\zeta_i)^{j-1}J_{j-1}(\zeta_i) ~~~~~~~~~~ 
{\mbox{\rm for}}~~i = 1,2 \cr
B_{ij} &=& (-\mu_{i-2})^{j-1}I_{j-1}(\mu_{i-2}) ~~~~~~~~~~ 
{\mbox{\rm for}}~~ 3 \leq i \leq N_f+2 ~,
\end{eqnarray}
and the $N_f\!\times\! N_f$ matrix $A$ is as in (\ref{Aconv}). To find
the corresponding double-microscopic spectral density, we make use
of the Bessel relation
\beq
\frac{d}{dx}\left[x^nJ_{n}(x)\right] = x^nJ_{n-1}(x)
\eeq
to get
\beq
\rho_S^{(N_{f})}(\zeta;\mu_1,\ldots,\mu_{N_{f}}) =
C~\frac{(-1)^{[N_f/2]+1}|\zeta|}{2
\prod_f(\zeta^2+\mu_f^2)}~\frac{\det \tilde{B}}{\det A} ~,\label{rhoS}
\eeq
where the  $(N_f\!+\!2)\!\times\!(N_f\!+\!2)$ matrix $\tilde{B}$ is defined
by
\beq
\tilde{B}_{1j} ~=~ (\zeta)^{j-2}J_{j-2}(\zeta) 
\eeq
and $\tilde{B}_{ij}\!=\! B_{ij}$ for $i\!\neq\! 1$. The general $n$-point
correlators follow from eqs. (\ref{correl}) and (\ref{KS}). It finally
remains to fix the constant $C$. We do this as before by the matching
condition $\rho_S^{(N_{f})}(\zeta\!\to\!\infty,\mu_1,\ldots,\mu_{N_{f}}) 
\!=\! 1/\pi$. This gives 
\beq
C ~=~\! (-1)^{[N_f/2]} ~.
\eeq
Substituting this into eqs. (\ref{KS}) and (\ref{rhoS}), the results
agree with what has recently been obtained by an
explicit computation in random matrix theory \cite{DN}.
As mentioned earlier, the case $\nu\!\neq\! 0$ 
can be extracted from the general formula for $\nu\!=\!
0$ by setting $\nu$ fermion masses equal to zero in a theory of $N_f\!+\!
\nu$ fermions. 

\noi
The case of the ordinary unitary ensemble, conjectured relevant for
QCD with an even number of flavors in $(2\!+\!1)$ dimensions \cite{VZ,DN1}, 
can be treated in entirely the same fashion now that the finite-volume
partition function is known explicitly \cite{DN1}. There are also
analogous relations between finite-volume partition functions and
(double-) microscopic spectral correlators for the cases corresponding
to orthogonal and symplectic random matrix ensembles.

\noi
We end by a few comments on how these results can be intuitively understood.
As we have seen, the (double-) microscopic correlators for the theory
with $N_f$ fermions can not be computed from the associated finite-volume
partition function for $N_f$ fermions alone. We need also to know the
finite-volume partition function for two more flavors. These additional
fermions, of imaginary mass\footnote{There is no ambiguity associated
with the analytic continuation to imaginary mass here. The finite-volume
partition functions are explicitly given in terms of Bessel functions, 
defined in the whole complex plane.}, act as sources that turn the
partition function into a generating function. It is not surprising
that the kernel, representing two degrees of freedom need two such
additional fermions to probe the dynamical distribution of Dirac 
eigenvalues. To close full circle, we still need to understand how the
master formula (\ref{mf}) can be derived directly from QCD alone.


\end{document}